\documentstyle[epsf,aps,prl,multicol]{revtex}

\newlength{\framewidth}
\newlength{\framewidthtwocolumn}
\framewidth=\textwidth
\framewidthtwocolumn=\textwidth
\advance\framewidth by -1.6ex
\advance \framewidthtwocolumn by -\columnsep
\divide\framewidthtwocolumn  by 2
\advance\framewidthtwocolumn by -1.6ex
\newlength{\minipagewidth}
\minipagewidth=\framewidthtwocolumn
\advance\minipagewidth by -5em

\def\half{{\scriptstyle{\frac{1}{2}}}}
\def\la{{\ \leftarrow\ }}
\def\Sd{\hat{S}}

\def\bxi{{\text{\boldmath\mbox{$\xi$}}}}

\def\bzero{\text{\bf 0}}

\def\x{{\text{\boldmath\mbox{${x}$}}}}

\def\n{{\text{\boldmath\mbox{${n}$}}}}

\newlength\mycaptionwidth
\mycaptionwidth=\textwidth
\advance\mycaptionwidth by -5 cm

\def\rnd{\,\text{\rm Rnd\index{Zufallszahlen}}\,}

\def\erf{\,\text{{\rm erf}}\,}

\def\ld{\,\text{\rm ld }\,}

\newcounter{stepcounter}
\newcounter{algocounter}

\def\TEXT#1{{\left\{
\text{\rm
\begin{minipage}{\minipagewidth}
{#1}
\end{minipage}}
\right\}
}}

\def\mysmallpicture{\leavevmode\unitlength
5mm\epsfverbosetrue\epsfxsize=176pt\epsfysize=151.5pt\begin{picture}(12,12)(-1.8,-1.7275)\put(-1.8,-0.7275)}
\def\closepicture{\end{picture}}

\def\Schroedi{{\sc Schr\"o\-din\-ger}}

\def\Hamilton{{\sc Hamilton}}

\def\Coulomb{{\sc Coulomb}}

\def\Sommer{{\sc Sommerfeld}}

\author{Thomas Fricke \footnote{\rm Email:  thomas@summa.physik.hu-berlin.de}}
\title{Quantum Mechanics Simulated as Branching Process}
\address{
  Institut f\"ur Physik\\
  Humboldt--Universit\"at\\
  Invalidenstr.~110\\
  10\,115 Berlin}
\begin{document}
\maketitle
\begin{abstract}
  Diffusion processes with branching play an important role in statistical
dynamics.
  They are a common approach to the  computing of quantum mechanical
groundstates,
  \cite{feynman,creufree,Reynolds:al,cepal} and  serve as models for population
  dynamics and as physical pictures for biological evolution,
  \cite{ebfe}.

  On a computer the efficiency of this  simulation method is
  limited by the approach to the infinitesimal time step, which is necessary to
  perform alternating diffusion and branching steps.

  In this paper, a method is described, which eliminates the
  infinitesimal time step for a certain class of branching processes,
  if the process of interest can be ``embedded''
  into another process, which is solvable by other  analytic
  and/or numerical methods.
  The simplest choice for the embbeding process is given by a process
  with a constant branching rate, which dominates the rate of the
  embedded process.

\end{abstract}
\begin{multicols}{2}
\section{Introduction}
A population of random--walkers $\phi(\x,t)$ which independently move,
multiply and die shall be described by an equation of motion where the
movement is given by a diffusion process, while the branching process
depends on the spatial coordinate $\x$. The probability for either a birth or a
death
process occuring during an infinitesimal  time--intervall $[t,t+dt]$
in the spatial interval $[\x,\x+d\x]$
is given by $|S(\x)| \phi(\x,t) \,d^d\x\,dt$, where $S(\x)<0$ denotes decay
processes and $S(\x)>0$ denotes birth processes.
\begin{eqnarray}
  \Pr&\TEXT{a birth process occurs during $[t,t+dt]$ in the spatial
    interval $[\x,\x+d\x]$}\\ &=\left\{
  \begin{array}[h]{rl}
    0,&\text{if\ }S(\x)<0,\\
    S(\x)\phi (\x,t)\,dt\,d^d\x,&\text{if\ }S(\x)>0, \\
  \end{array}\right.\\
  \Pr&\TEXT{a decay process occurs during $[t,t+dt]$ in the spatial
    interval $[\x,\x+d\x]$}\\ &=\left\{
  \begin{array}[h]{rl}
    -S(\x)  \phi (\x,t)\,dt\,d^d\x,&\text{if\ }S(\x)<0 ,\\
    0,&\text{if\ }S(\x)>0.\\
  \end{array}\right.
\end{eqnarray}
To stress the close relation to
imaginary time quantum--mechanics  the reproduction operator
$S$ is introduced
as the negative potential $S(\x)=-V(\x)$.
The simplest case, where the the diffusion is homogeneous and
isotropic the dynamics is governed by the equation
\begin{equation}
  \label{IMS}
\partial_t \phi(\x,t) =
\left\{\half {\partial^2_{\x} } - V(\x)  \right\} \phi (\x,t)
\end{equation}
This equation is equivalent to an imaginary time \Schroedi--equation,
with the  dynamics
\begin{equation}
  -H=\half \partial^2_{\x} - V(\x),
  \qquad \partial_t \phi  = -H \phi.
\end{equation}
The formal analogy between imaginary time \Schroedi--equation and
statistical physics has been studied since  \cite{feynman},
a contemporary approach is given in \cite{roep}.
Most of the works focuses on the relation between the imaginary time
\Schroedi--equation and equilibrium statistics.
However, this paper emphasises, that there is a more natural interpretation
as a non--equilibrium process, which has been explicitely stated in
\cite{ebfe} and which is implicitely used in \cite{Reynolds:al,cepal}.
\section{Theoretical foundation for the imaginary time \Schroedi--equation}
In this section, the method to measure the energy of the quantum--mechanical
groundstate and to stabilize the population shall be derived.
In contrast to  the common bilinear product $ \langle\phi | H | \phi \rangle $,
a flux through the population shall  be used
to measure the energy.
The population is  stabilized on  the average, by subtracting
the groundstate energy $E_0$, which is determined ``on the run''.
\subsection{Formal solution}
In analogy to real time  \Schroedi--equation
$-i\hbar \phi = H \phi$ equation (\ref{IMS}) can be solved
using the same eigen--function system $ \psi_n$  of $H$ with
$H\psi_n=E_n \psi$.
This way one obtains exponentially growing (decaying) contributions
for $E_i<0$ ($E_i>0$), instead of oscillating ones
\begin{equation}
  \phi (\x,t) =\sum_n c_n e^{ -E_i t } \psi_n(\x), \qquad
\label{cn}  c_n=\langle \phi_0 | \psi_n\rangle .
\end{equation}
Of course, the groundstate $\psi_n $ grows faster (or decays
slower) than all other states and will dominate the ensemble
for $t \to \infty $, thus
on the long run the normalised population converges to
the quantum--mechanical groundstate
\begin{equation}
 \label{groundstate}
  \psi_0(\x)=\lim_{t\to \infty } {\phi (\x,t)}\left/{\int \phi (\x',t) d
\x'}\right..
\end{equation}
\subsection{Measuring the energy}
It is remarkable, that the energy of $ \phi $ is
related to a flux through the population,
which is defined on an area $G$ by
\begin{equation}
  \label{flux}
  F(t) =\int_G H \phi \,d\x
  = -\int_{\partial G} \nabla \phi  \,d\n +\int_G V \phi
\, d^d\x.
\end{equation}
Obviously, an exploding or decaying system cannot be in equilibrium.
Of course, even if the total population is stable $\int_G \phi
(\x,t)=$constant,
there are areas, where
the random--walkers are born, and other areas reached by diffusion,
where they die.
Thus, the RDS cannot be  related to an equilibrium system.
However, under certain conditions it is convenient to have
a stationary flux, which is equivalent  to a system with a stable population.
\subsection{Stabilizing  the population}
\label{stabil}
The main idea to stabilize the population is to determine the energy
of the groundstate and to subtract it from the effective dynamics.
This has the effect of putting the lowest eigen--value
to zero by hand.
As $\phi $ is the formal solution of
\begin{equation}
  \phi(t)=\exp \left( -Ht \right)  \phi(0),
\end{equation}
it is useful to notice, that for an arbitrary time--depending
function $\bar E(t)$ the equation
\begin{equation}
{\partial_t \phi(\x,t)}= \left\{  \bar E(t) -H \right\} \phi(\x,t),
\end{equation}
with the same  $c_n$ as defined  in equation (\ref{cn}), is solved by
\begin{eqnarray}
\phi (\x,t) =\sum_n c_n e^{\int_0^t E(t')\,dt' -E_i t } \psi_n(\x).
\end{eqnarray}
Thus, $ \phi $ can also be stabilised by subtracting
an average energy with the property
$
  \int_0^t \bar E(t')\, dt' \cong E_0 t, $
 for $t \to \infty $,
which is automatically implied if
$
  \lim_{t \to \infty }\bar E(t) = E_0.
$
\ Therefore,  the average
\begin{equation}
\bar E (t)={\int_0^t F(t') \,dt'}\left/{\int_0^t \int_G \phi(\x',t') \,
d\x'\,dt'}\right.,
\end{equation}
is  chosen converging $\bar E(t) \to E_0 $ because $\phi\to \psi_0$ for $t \to
\infty $
by   (\ref{groundstate}) and inserting $\phi$   into   definition (\ref{flux}).
\section{Simulation without infinitesimal time--steps}
The starting point is the common time--step algorithm.  The
simulation needs $N$ time--steps to evolve from time $t=0$ to
the final time $T$,  each step of length $ \varepsilon = T/N$.
With $\rnd$ and $\bxi$ denoting standard uniform and normal
random--numbers the simplest formulation is:\\[1ex]
\parsep-1ex
for $n\la 0 $ to $N$
  \begin{enumerate}
    \vspace*{-1ex}
  \item $t\la{n \varepsilon  }$,
        \vspace*{-1ex}
  \item move $\x_r \la \x_r +\sqrt{\varepsilon } \xi $,
        \vspace*{-1ex}
  \item if $ |S(\x_r)|\varepsilon > \rnd $ then\\
     \hspace{1em}perform a birth ($S>0$)  or death ($S<0$) process,
     \label{step}
         \vspace*{-1ex}
  \item next $n$.
  \end{enumerate}
At first, it is assumed,
that there is an upper bound $\hat S> S(\x)$ allowing to split step \ref{step}
into
two parts:
\begin{itemize}
  \item[3a.] if $ |\hat S|\varepsilon < \rnd $ reject any branching, next $n$,
             \vspace*{-1ex}
  \item[3b.] if $ {|S(\x_r)|} < \rnd{\hat S} $ then \\
     \hspace{1em}perform a birth ($S>0$)  or death ($S<0$) process.
\end{itemize}
The probability of a branching event is the probability $p_a=|\hat
S|\varepsilon$
to pass step 3a times  the probability $p_b=|S(\x_r)|\left/\hat S \right. $ to
pass step 3b without rejection.
The result is the correct branching rate $p_a p_b=|S(\x_r)| \varepsilon $.

Now there is an inner loop, 1--3a, and an outer loop, 1--3b.  In
the inner loop there is no need to evaluate the exact branching
potential $S(\x)$.  The probability, {\it not to leave} the inner loop
after one step is $Q^1=(1-\hat S \varepsilon )$.  The probability, not
to leave the inner loop after $n$ steps, or equivalent until time
 $t=n\varepsilon $ is given by $Q^n=(1-\hat S \varepsilon)^n=q_{\varepsilon
}(t)$.
 The limit $ \varepsilon \to 0$ can be
evaluated
\begin{equation}
  \label{i:rejections}
  q_{\varepsilon}(t) =
  %
  (1-\hat S \varepsilon )^{t/\varepsilon }
  \stackrel{\varepsilon \to 0}{\longrightarrow}
\left(e^ {-\hat S} \right)^t=e^{ - \hat St}.
\end{equation}
Thus the whole loop can be replaced by:\\[1ex]
while $t<T$
  \begin{enumerate}
  \item draw an exponentially distributed random--number \\
    $ \tau \la \log    \left(1-\rnd \right)/{\hat S}, \quad
    t\la t +\tau  $
    \vspace*{-1ex}
  \item move $\x_r \la \x_r +\sqrt{\tau } \xi $,
    \vspace*{-1ex}
  \item if $ {|S(\x_r)|} < \rnd{\hat S} $ then \\
     \hspace{1em}perform a birth ($S>0$)  or death ($S<0$) process,
     \vspace*{-1ex}
   \item next exponential time step.
  \end{enumerate}

This way, the time $ \tau $ between to proposed branchings is
related to the upper bound $\hat S$ for the potential $|S(\x)|$, if
the branching has been succeeded or rejected.  The rate for
evaluating the true $ S(\x)$ is $ {\hat S} $, which can be chosen to
be of a typical order of magnitude for the given problem.  According
to the considerations to stabilise the groundstate,  section
(\ref{stabil}), we choose $S(\x,t)=\bar E(t)-V(\x)$.

Although, the proof of the algorithm has not been given for time
dependent $S(\x,t)$, this generalisation is obvious.
Furthermore,
in the long time limit, the fluctuations of $\bar E(t) $ fade away.
\section{The representation of the wave function}
The simulations have been performed by ensembles of
$R$ idependent random--walkers, each described by two variables, the
coordinate $\bxi_n^r$ and the time $\theta_n^r$, where and when the
 $r$--th walker has been a ``candidate'' for its $n$--th branching
process, independent of a branching has been performed or not.
Since the last branching proposal at time  at location
the random--walker has done a free
random--walk. When it is considered to branch again at time $t$, its
new position is given by a normal distributed random--number with mean
$\bxi^r_n$ and variance $ \sigma =\sqrt{t-\theta^r_n}$. With
$
  g(\x,\sigma^2 )={ \left( 2 \pi \sigma^2 \right)^{-d/2} }
  \exp \left( - {\x^2}/{2  \sigma^2 }\right)
$
the wave--function is represented by
\begin{equation}
  \phi(\x,t)=\sum_{r=1}^R g(\x-\bxi_r,t-\theta_r)
\end{equation}
and averaging the flux (\ref{flux}) reduces to the computation and summation
of
integrals
\begin{eqnarray}
\nonumber
  \int_0^t F(t')\,dt'&=&
\sum_{r=1}^R\sum_{{ n  = 1}\atop{ \theta_{n}^r < t}}
 U(\theta_{n}^r-\theta_{n-1}^r,\bxi_{n-1}^r)\\
  \text{with }
 U(\tau,\bxi)&=&
 \int\limits_{0}^{\tau } d t'
  \int d^d\x\, V(\x)\, g \left(\x-\bxi,t'\right).\label{U:define}
\end{eqnarray}
\section{Examples}
The quality of the simulations depends on the computation of
$U(\tau ,\x)$. For all standard potentials
like polynomials and  \Coulomb--like
potentials, including interacting electronic systems,
$U$ can be solved analytically.
\subsection{Polynomials and the harmonic oscillator}
In (\ref{U:define}) the function  $U(\tau,\x )$ has been defined
as the time average of the energy.
For an arbitrary polynomial
\mbox{$
P(\x)=\sum_{k_1,\dots,k_d}
q_{k_1,\dots,k_d}\prod_{i=1}^d x_i^{k_i}
$}
one finds
\begin{equation}
  \label{Wpoly}
  U(t,\x)=\sum_{k_1,\dots,k_d}
  q_{k_1,\dots,k_d}\prod_{i=1}^d
  \sum_{j_i=0}^{k_i}\kappa_{j_i,k_i}\,
  t^{\frac{j_i}{2}+1}\,
  x_i^{k_i-j_i}
\end{equation}
with
$
\kappa_{j,k} = {{\frac{1+(-1)^{j}}{2}}}\,
{{k}\choose{j}}\,
{-{1}/{2}\choose{j}/{2}}\cdot\,
\frac{(-2)^{{j}/{2}}}{j/2+1}
$
which is easy to implement.
For the $d$--dimensional  harmonic oscillator  $V(\x)=P(\x)=\half |\x|^2$,
one obtains the simple result
\label{WHarmosc}
\begin{equation}
U(t,\x)
=\half\, t\,|\x_0|^2+\frac{d}{4}\,t^2.
\end{equation}
\narrowtext
\begin{figure}[h]
  \begin{center}\vspace*{-3ex}
    \leavevmode
    \mysmallpicture{\epsfbox{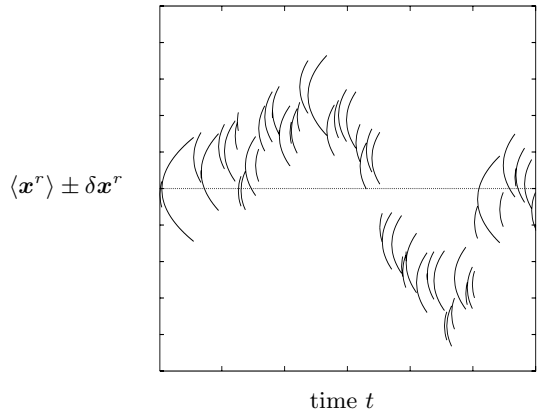}}
      \put(-4,4.7){\small $\langle  \x^r \rangle \pm \delta \x^r $}
      \put(4,-1){\small time $t$}
      \closepicture
   \end{center}
  \caption{paths of the harmonic oscillator for a single random--walker,
    $x^r \pm \protect\sqrt{t-\theta_n^r}$
    to demonstrate the variance of the normal distribution. The
    diffusion is free for a time $ \tau $ of the order of a typical
    time scale.}
  \label{fig:path:harmonic}
\end{figure}
\subsection{The \Coulomb--potential}
Although the \Coulomb\  $1/|\x| $ singularity has no upper
bound, the function $U$  can also be computated by analytical methods.

By scaling the \Hamilton ian gets $ H=-\half \partial^2_{\x}   +{\alpha}/{|\x|}
$
with the  \Sommer--constant $ \alpha $ being the only free
parameter, thus  $ E_0=-\half \alpha^2$.
The integrals (\ref{U:define}) are solved in polar coordinates
performing the integrations in the order $\varphi\to\theta\to r\to t$, with the
angles defined according to the axis $\bzero\to \x$.
The $U_{\sf H}$--function  reads
\begin{equation}
\label{Coulomb:Wirkung}
U_{\sf H}(t,\x)=\alpha\sqrt{2t}\ \ u\left({{|\x|}}\left/{ \sqrt{2t}}
\right.\right),\\
\end{equation}
with $u(x')=
\left({1}/{(2x')}+x'\right)\erf(x')+e^{-{x'}^2 }/\pi-x'$.
The function $u(x')$ has a very simple structure,
and is tabulated at the beginning of each run.
The integrals evaluated for the computation of the
\Coulomb--potential may easily be generalised to simulate
the ortho--helium atom.
The \Hamilton ian
\begin{equation}
  H=-\sum_{i=1}^2\left(\half\partial^2_{\x_i} -{2\alpha}/{|\x_i|} \right)
  +  {\alpha}/{|\x_1-\x_2|}
\end{equation}
allows a separation of the $6+1$--dimensional integration.
For  $V_i= 2 \alpha / |\x_i| $
there is  one integration for $\x_i$ and another one
for the coordinate which $V_i$ is independent of.
The $V_{12}= \alpha /|\x_1-\x_2| $ integration may be transformed into the
centre of
mass system, resulting in a constant factor and another integration
for the relative coordinates which reduces to a \Coulomb--integration
again.
The final result reads
\begin{eqnarray}
  U_{\sf{ He}}(t,(\x_1,\x_2))=2\sum_{i=1}^{2}U_{\sf{ H}}(t,\x_i)+U_{\sf
H}(t,\x_1-\x_2).
  \label{Helium:Wirkung}
\end{eqnarray}
\vspace*{-2ex}
\section{The quality of the results}
\vspace*{-2ex}
To keep the number of random--walkers constant,
every time a random--walker multiplies (dies) another one
dies (multiplies) with probability $1/(R-1)$.
For this ``antagonist'' the actualization of the coordinates has to
be carried out, too.
This way, $R$ does no fluctuate any more, otherwise the fluctuations
of $R$ would cause a diffusion process of $R$, which is avoided
for convenience.
\begin{figure}[h]
  \begin{center}\vspace*{-3ex}
    \leavevmode
    \mysmallpicture{\epsfbox{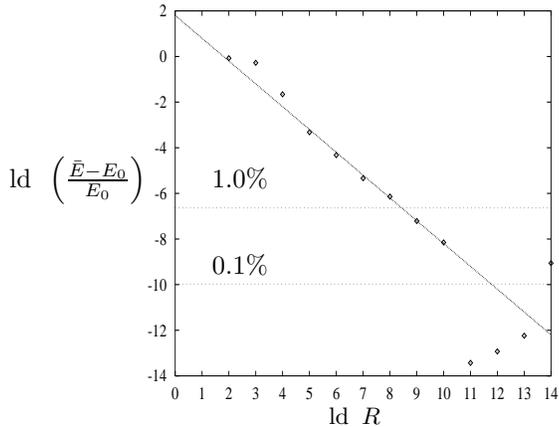}}
      \put(-4.5,5){$\ld \left(\frac{\bar E-E_0}{E_0} \right)$}
      \put(4,-1.3){$\small\ld R$}
      \put(1,5){1.0\%}
      \put(1,2.7){0.1\%}
      \closepicture
  \end{center}
  \caption{Systematic error of $E_0$ for the hydrogen atom, depending on the
number of random--walkers.
    The straight line represents $1/R$.}
  \label{fig:error:R}
\end{figure}
The antagonist rule
introduces an unknown additional coupling among two walkers considered as
independent.
This results in a $R^{-1}$ law, the large number of $R-1$ possible
partners weakens the individual coupling.
Thus the order of the systematic  error is higher than the statistical error
of $R^{-1/2}$ and can be neglected, if $R$ is sufficiently large (Fig.
\ref{fig:error:R}).
Typical values are  $0.1 $ percentage error for $R=2^{10}=1024$.

\begin{figure}[h]
  \begin{center}\vspace*{-3ex}
    \leavevmode
    \mysmallpicture{\epsfbox{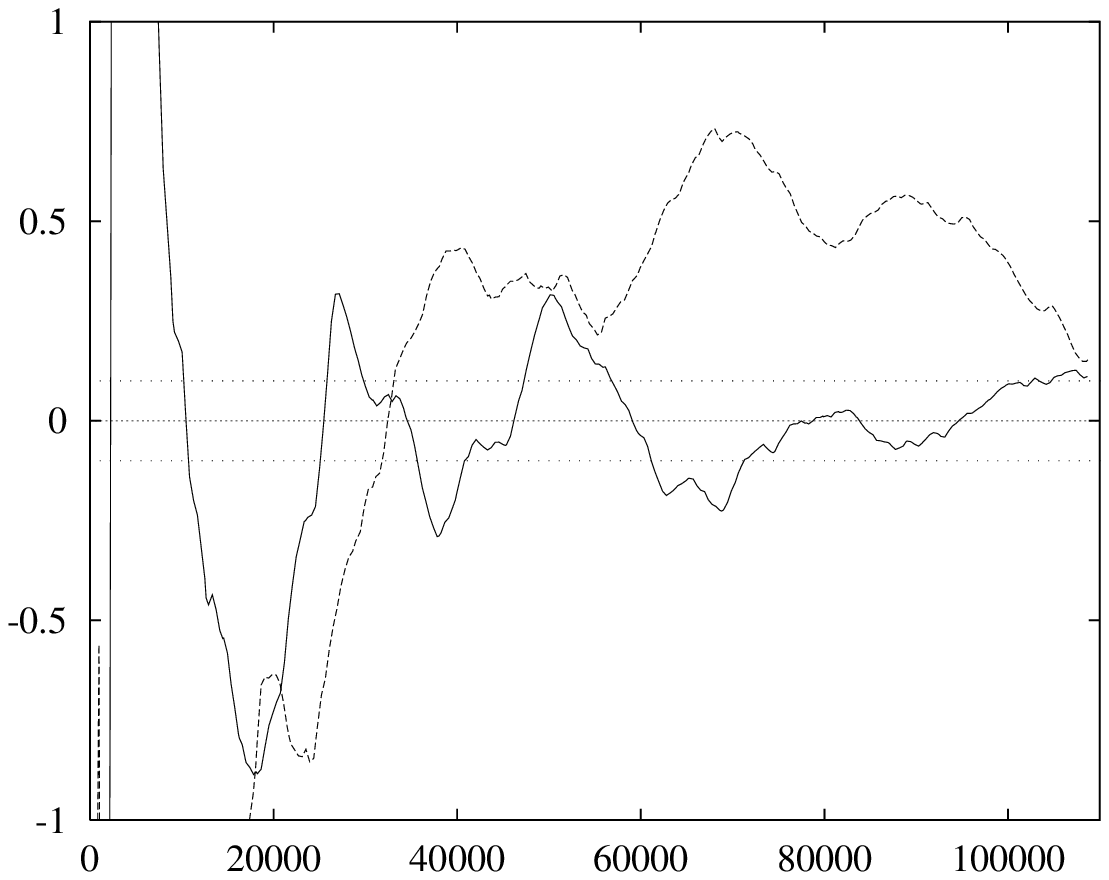}}
      \put(-4,5){$ \frac{\bar E-E_0}{E_0}/$\%}
      \put(4,-1.3){\small CPU time}
      \closepicture
  \end{center}
  \caption{$E_0$ for typical run of a helium simulation, computing time
vs.~percentage error}
  \label{fig:helium:run}
\end{figure}

For harmonic or {\Coulomb}\ potentials using a constant $\Sd(\x,t)$  the
condition
$\Sd > S(\x)$ is violated.
However, due to the fast decay of the wave function
this error decays exponentially
fast for the harmonic oscillator.
For the \Coulomb\ like problems it only affects a neglectible
area.
Fig.~\ref{fig:helium:run} gives an impression of two typical runs for the
helium
problem approaching the experimental groundstate energy better than $0.1\%$.
\vspace*{-2ex}
\section{Conclusion}
\vspace*{-2ex}
The algorithm has proved its capability to simulate
imaginary time \Schroedi--equations very efficient and accurate.  The
method of embedding has been extended to all problems, which can be
formulated as branching processes, and it has been generalised
to arbitrary time depending rates $ S(\x,t)$ \cite{fricke}.
Without respecting the bosonic or fermionic properties of the wave
function, in multi particle simulations there is a mixture of
symmetric and anti--symmetric components.  Because the bosonic
wave--function has a lower energy than the fermionic one, for long
times the solution converges to the bosonic groundstate.  Thus, the
algorithm in its present implementation is ideal for the simulation of
interacting bosons \cite{roep}.

One interesting property of the algorithm for quantum--mechanics
is the fact, that \Coulomb\  2--particle integrals can be solved
analytically.
 Its importance is limited by the question, if it will be
possible to extend it to fermionic systems.  For all kinds of
\Coulomb--systems an identity like (\ref{Helium:Wirkung}) holds as
well and the singularities may be integrated and smoothed out.  At
least, it should be possible to introduce the nodes of a related
problem as absorbing boundaries similar to  \cite{cepal}.

However, the main advantage is opposite to \cite{Reynolds:al,cepal}
that there is no need of a ``guiding wave function''.
 Therefore, the algorithm is easier to implement.
The related ideas concerning fermionic systems \cite{cepal}
and excited states \cite{cep:bernu},
could be applied as well.
 
\end{multicols}
\end{document}